\documentclass{elsarticle}
\usepackage[english]{babel}
\usepackage{xcolor}
\usepackage{ulem} 
\usepackage{amssymb,amsmath}
\usepackage{graphicx,subfigure}
\usepackage{hyperref}

\DeclareMathOperator{\divv}{div}

\newcommand{\eqs}[1]{Eq.~\eqref{#1}}
\newcommand{\f}{\frac}
\newcommand{\m}{\mathbf}
\newcommand{\bs}{\boldsymbol}

\graphicspath{{D://localtexmf//Mytex//flocking//articles//2013//figures//}}
\begin{document}
\title{The hydrodynamic description
for the system of self-propelled particles: Ideal Viscek fluid.}
\author[ods,nce]{Oleksandr Chepizhko}
\author[ods]{Vladimir Kulinskii}

\address[ods]{Department for Theoretical
Physics, Odessa National University, Dvoryanskaya 2, 65026 Odessa,
Ukraine}
\address[nce]{Laboratoire J.A. Dieudonn\'{e}, Universit\'{e} de Nice Sophia Antipolis, UMR 7351 CNRS , Parc Valrose, F-06108 Nice Cedex 02, France}
%
%
\begin{abstract}
We use the method of the microscopic phase density to get the kinetic equation for the system of self-propelled particles with Vicsek-like alignment rule. The hydrodynamic equations are derived for the ordered phase taking into account the mean-field force only. The equation for the hydrodynamic velocity plays the role of the Euler equation for the self-propelled Vicsek fluid. The hydrodynamics of such ideal self-propelled fluid demonstrates the dynamical transition from disordered initial state to the completely ordered motion.
To take the noise into account we consider how the framework of the local equilibrium approximation affects the hydrodynamic equations and the viscous tensor and show that in such approximation the shear viscosity vanishes.
\end{abstract}
\maketitle
\section*{Introduction}
The study of collective phenomena in the systems of active agents is an important and fast developing field of statistical physics due to the potential  application to the dynamics of living matter \cite{socialdynamics_rmp2009,spp_vicsek_physrep2012}. This phenomenon is widely observed in nature at different levels of organization including the synchronicity in insects behavior, bird flocking, traffic and complex social behavior \cite{socialdynamics_rmp2009,spp_helbingtraf_rmp2001,spp_ramaswami_annurev-conmatphys2010,spp_insects_scireports2013}. These systems are essentially non-equilibrium and their dynamics is not strictly Hamiltonian due to the information exchange which includes in general not only the positions of the neighbors but also their velocities.

There are two classes of models that are widely used. Here we adopt the terminology of the works \cite{spp_nonholon_coopcntrl2005, spp_cukersmale_ieee2007}. The first class can be called ``dynamic''. In the models of this class the energy exchange between moving agents and an external source (energy depot) takes place \cite{book_scweitzeractiveagents} (for the recent review see also \cite{spp_activebrownebeling_epj2012}). This class is well suited for the description of self-propelled microorganisms immersed in an external fluid, which causes the additional hydrodynamic interaction between agents. The driving forces are mostly due to the gradients of the external factors meaningful for the living organisms such as light, food concentration etc. \cite{spp_microorg_advphys2000}.

The second class is formed by the ``spin-like'' models of self-propelling particles (SPP) since the speed of particles is constrained to be constant. Thus the velocity plays the role of  spin in lattice models of statistical physics. These models are essentially non-holonomic because of the nontrivial control of the angular velocity of a particle rather than its speed. The models of this class are aimed for the description of the cooperative behavior due to the exchange of information between the agents \cite{spp_reynolds_compgraph1987}. The most prominent and minimal model of this class was introduced
by T.~Vicsek and collaborators in \cite{spp_cva_prl1995}. Due to its simple formulation yet nontrivial character it plays the role analogous to the Ising model in the theory of phase transitions. We call it the Standard Vicsek Model (SVM) \cite{spp_vicsek_physrep2012}. It has generated a series of publications \cite{spp_nonholon_coopcntrl2005,spp_cukersmale_ieee2007,spp_tonertu_prl1995, spp_aldanahuepe_network_jsp2003, spp_gregchate_prl2004,spp_clustrods_pre2006} because of its simplicity and rich behavior.
Recently, a promising approach which unifies these two classes has been put forward in \cite{spp_activecolloid_nature2013} by the realization of the Viscek type of interaction via dipole-dipole potential in motile colloids.

There are many questions that still need to be clarified for the model. The Vicsek model can be modified to include nematic alignment \cite{spp_clustrods_pre2006,spp_rods_prl2010}, non-metric interactions \cite{spp_kineticihle_pre2012}. Besides, it can be extended by the inclusion of obstacles as the heterogeneous environment \cite{spp_obstaclesperuanichip_prl2013,PhysRevLett.111.160604}.   The type of phase transition is one of the main open problems. It is not known exactly either it is continuous \cite{spp_vicseknagypha2007,spp_aldanahuepe_prl2007, spp_aldana_jmpb2009,spp_albano1order_pre2009} or discontinuous \cite{spp_gregchate_prl2004, spp_gregchaiteanswer2mex_2007,spp_chaitesumul_pre2008, spp_antialign_jop2011} and whether this depends on the type of the stochastic perturbation.
The mean-field approximations of similar network model \cite{spp_aldanahuepe_prl2007} and of the Vicsek model itself \cite{spp_kineticus_proc2009} demonstrate the dependence of the character (continuous or discontinuous) of the transition on the type of noise. Such feature is known for another seminal model of synchronization - the Kuramoto model  \cite{spp_kuramoto1975,spp_kuramotomodel_rmp2005}.
The SVM can be thought as the dynamic version of the Kuramoto model and at least in the mean-field approximation shares qualitatively the same dependence of the transition (sub- or super-critical) on the type of the noise \cite{spp_vicsekuramotous_physa2010}.

Approaching this problem theoretically leads to the formulation of the proper kinetic equation which adequately reflects the basic physical mechanism of self-organization and justifies the hydrodynamic equations proposed previously from phenomenological reasoning \cite{spp_tonertu_prl1995, spp_tonertu_pre1998,spp_us_eurphyslet2005}. The first attempt to do this was launched by E.~Bertin et al. \cite{spp_gregorie_kinetic_pre2006,spp_bertingregoire_jphysa2009} and  T.~Ihle \cite{spp_kinetichydro_pre2011, spp_kineticihle_pre2012}. In Bertin's approach the kinetic equation was obtained following standard Boltzmann derivation using two-particle collision integral. This gave the explicit expressions for the coefficients in the hydrodynamic equations.
This approach uses disordered state as the zero-order approximation to construct the distribution function. It shows good agreement with numerical results close to the transition point. Recently this approach was assessed in \cite{spp_boltzcritique_prl2013}, pointing at the troubles of the two-collision assumption.

In \cite{spp_kinetichydro_pre2011,spp_kineticihle_pre2012} the Liouville formalism was used as the starting point. The factorization of exact $N$-particle distribution function to the product of one-particle functions lead to the kinetic equation for the one-particle function. The obtained kinetic equation took into account multi-particle collisions but neglected correlations between particles, which are very important in the SVM. Note that the shear viscosity term appeared in all approaches. But until now no experimental (numerical) evidence of the viscous shear has been reported for the SVM in bounded regions. In \cite{spp_gonnellashear_cejp2012} the influence of external shear flow has been studied for the SVM but it has no connection with the inner viscosity of the spp-fluid and can be considered as the specific case of the extrinsic noise \cite{spp_gregchate_prl2004, spp_aldananoiseswarming_pre2008}.

In this paper we use well-known method of the microscopic phase density \cite{book_klimontovich_statphys}. In the framework of this approach the kinetic equation is constructed directly on the basis of equations of motion without reference to $N$-particle distribution function. To the best of our knowledge firstly such approach for the self-propelling system with Vicsek-type of interaction was regularly used by P.~Degond and S.~Motsch in \cite{spp_degondmacrolimit_comptrend2007,spp_degondcontlimit_mmapp2008}.
In these works the hydrodynamic equations for the Vicsek fluid were proposed \cite{spp_degondcontin_mmmas2012,spp_degondcrowds_comphys2013}.  The method of microscopic phase density, on one hand, is well known from the statistical mechanics and was used to describe, for example, plasma, and on the other hand is different from the Boltzmann approach, or an approach that is based on the factorizing of the N-particle distribution function that are used in \cite{spp_gregorie_kinetic_pre2006,spp_bertingregoire_jphysa2009,spp_kinetichydro_pre2011, spp_kineticihle_pre2012}. Boltzmann approach, and factorization both assume that the system is diluted. The MPDF approach shows a way to take into account more complicated correlations. In this paper we consider only the simplest case, deriving an equation analogous to the Euler equation for the usual fluid. The question of the ideal limit of the Vicsek fluid has never been posed before. We show that such limit is an important starting point for the hydrodynamics of the self-propelled fluid.



We use the basic equation for the microscopic phase density functional as the starting point because of its direct connection with the equation of motion. Our prime interest is to get an equation analogous to the Euler equation, but for the self-propelled fluid from the basic equations of motion, and  to compare it with known results: phenomenological as well as derived from kinetic approaches.

The paper is organized as follows. In Sec.~\ref{sec_mfpd} we derive the equation for the microscopic phase density functional of the SVM which leads to the formal kinetic equation with the corresponding collision terms. Then we obtain the hydrodynamical equations. In Sec.~\ref{sec_idealhydro} we consider the hydrodynamical limit of the ideal self-propelled Vicsek-like fluid. By the ideal fluid we understand a fluid where the correlations are neglected though the particles interact with each other via the self-consistent field.
In the Sec.~\ref{sec_visc} we consider local equilibrium approximation as the simplest way to take noise into account. We show how it changes the hydrodynamic equations, and compare our coefficients with the ones from \cite{spp_bertingregoire_jphysa2009}.
We discuss the question about the existence of the shear viscosity for such fluid. There we study the viscosity of this fluid from the theoretical point of view using the equations derived in previous sections. We postulate the problem of numerical simulations of Couette flow for Vicsek-like model and discuss preliminary results. The analysis of the results and problems for future studies are given in the concluding section.

\section{The equation for the Microscopic Phase Density
Functional}\label{sec_mfpd}
The standard tool for the derivation of the
hydrodynamic as well as kinetic equations for dynamical systems is
the method of microscopic phase density functional (MPDF)
\cite{book_klimontovich_statphys}: %
\begin{equation}\label{micro_phasedensd}
\mathcal{N}(x,t) = \sum_i \delta(x-x_{i}(t))
\end{equation}
where $x = (\mathbf{r}, \mathbf{v})$. It obeys the conservation
law:
\begin{equation}\label{balance_micro_ph_d_new}
 \partial_{t}\,\mathcal{N} +\mathbf{v}\,\partial_{\mathbf{r}}\,
\mathcal{N}
+ \partial_{\mathbf{v}}
\left(\,\dot{\mathbf{v}}\,\mathcal{N}\,\right)=0\,.
\end{equation}
The corresponding number density and the density flow are:
\begin{eqnarray}
  \rho^{(m)}(\mathbf{r},t) &= \int
\mathcal{N}(x,t)\,d\mathbf{v}\,, \label{dens}\\
  \mathbf{j}^{(m)}(\mathbf{r},t) &=  \int
\mathbf{v}\,\mathcal{N}(x,t)\,d\mathbf{v}\,,\label{flow}
\end{eqnarray}
where the superscript ``$m$'' stands for the ``microscopic''. We
take the equation of motion for the Vicsek model \cite{spp_cva_prl1995} without the noise term as follows:
\begin{equation}\label{eom}
	\frac{d \mathbf{v}_{i}}{d t} = \boldsymbol{\omega}_{i}\times
\mathbf{v}_{i}\,\,,
\end{equation}
which describes the fact that the kinetic energy of a particle
is conserved. The original Vicsek algorithm \cite{spp_cva_prl1995}
is based on the alignment of the particle velocity to the average
direction of the neighbors:
\begin{equation}\label{def_avelocity0}
\mathbf{w}_i = \frac{\sum_j \mathbf{v}_j
K(\mathbf{r}_i-\mathbf{r}_j)}{|\sum_j \mathbf{v}_j
K(\mathbf{r}_i-\mathbf{r}_j)|}\,\,,
\end{equation}
where $K(\mathbf{r}_i-\mathbf{r}_j)$ is the ``microsopic''
summation kernel and therefore: %
\begin{equation}\label{eq_angvel}
  \bs{\omega}_{i} = \gamma\,\m{v}_{i}\times \m{w}_{i}\,.
\end{equation}
Usually, the Heaviside step function is used for this purpose
with the characteristic region of several average distance between the particles.

To avoid the difficulties in subsequent derivation connected with
the non-additive structure of Eq.~\eqref{def_avelocity0} we
use another definition for $\m{w}_{i}$:
\begin{equation}\label{def_avelocity}
\mathbf{w}_i = \sum_j \mathbf{v}_j
K(\mathbf{r}_i-\mathbf{r}_j)\,,
\end{equation}
Otherwise using of Eq.~\eqref{def_avelocity0} leads to the essential multi-particle collision terms (see e.g. \cite{spp_kinetichydro_pre2011}). These two options differ by the scalar factor $\gamma$. Therefore one can expect that despite the difference in the microscopic equation of motion they lead in essential to the same macroscopic dynamics.

Because of its additive structure Eq.~\eqref{def_avelocity} can be expressed as the average of the MPDF:
\begin{equation}\label{eq_avelocityfield}
\mathbf{w}_i =
\int dx'\,K(\mathbf{r}-\mathbf{r}') \mathbf{v}'\mathcal{N}(x',t)
= \int
K(\mathbf{r}-\mathbf{r}')\,\mathbf{j}^{(m)}(\mathbf{r}',t)\,
d\mathbf{r}'\,\,.
\end{equation}
In such a case we get the equations similar
to the BBGKY hierarchy for molecular systems with pairwise interaction potential \cite{spp_degondcontin_mmmas2012}.

The equation \eqref{balance_micro_ph_d_new} takes the form:
\begin{equation}\label{micro_ph_d_full}
\partial_{t}\,\mathcal{N}(x,t)
 +\mathbf{v}\cdot\partial_{ \mathbf{r}}\, \mathcal{N}(x,t)
 +  \gamma \partial_{\mathbf{v}} \left(\int d x'\,
\left(\,\mathbf{v} \times \mathbf{v}' \,\right)\times\mathbf{v}\,
K(\mathbf{r}-\mathbf{r}') \mathcal{N}(x', t)\,\mathcal{N}(x,t)\right)=0\,\,.
\end{equation}
From this equation it follows that the equation for one-particle
distribution function $\overline{\mathcal{N}(x,t)} = n f_1 (x,t)$
is:
\begin{equation}\label{eq_balance_f1_microNN}
\partial_{t}f_1(x,t)
 + \mathbf{v}\cdot\partial_{\mathbf{r}} f_1(x,t)
+ \frac{\gamma}{n} \partial_{\mathbf{v}}\left(\int K(\mathbf{r} -
\mathbf{r}')\,
\left(\, \mathbf{v} \times \mathbf{v}'
\,\right)\times\mathbf{v}\,
\overline{\mathcal{N}(x,t) \mathcal{N}(x',t)}\, dx'\right)=0\,,
\end{equation}
where $\overline{ \mathcal{N}(x,t) \mathcal{N}(x',t)}$ is connected with
the pair distribution function $f_2$:
\begin{equation}\label{nn}
\overline{ \mathcal{N}(x,t) \mathcal{N}(x',t)} =
 n\,\delta(x-x')\,f_{1}(x,t)+n^2\,f_{2}(x,x',t)\,\,.
\end{equation}
Using the common definition of the correlation function $g_{2} =
f_{2} - f_1\,f_1$ we get:
\begin{equation} 	
\overline{\mathcal{N}(x,t) \mathcal{N} (x',t)} -
\overline{\mathcal{N}(x,t)}\, \,\overline{ \mathcal{N}(x',t)} =
n^2\,g_2(x,x',t) +n\,\delta(x-x')\,f_{1}(x,t)\,\,.
\label{fluctuations_of_pd}
\end{equation}
Thus
\eqs{eq_balance_f1_microNN} transforms into:
\begin{equation}\label{basiceq}
	\left(\partial_{t} +
\mathbf{v}\partial_{\mathbf{r}} +
\partial_{\mathbf{v}}\mathbf{F}
	\right)	f_1(x,t) =  I(x,t)\,\,,
\end{equation}
where $I(x,t)$ is the collision integral:
\begin{equation}\label{collisint}
I(x,t) = \gamma\,n\,\partial_{\mathbf{v}}\left(\, \int 	
\mathbf{v}\times\left(\mathbf{v}\times \mathbf{v}'\right)\,
K(\mathbf{r} - \mathbf{r'})\,g_2(x,x',t)\,dx' \,\right)\,\,.
\end{equation}
Also we introduce the mean force $\mathbf{F}$ due to the neighbors:
\begin{align}
	\mathbf{F} (x,t) & = n \gamma \int \mathbf{v}\times
\left(\, \mathbf{v}'\times \mathbf{v} \,\right) \,
K(\mathbf{r} - \mathbf{r'})\,f_{1}(x',t)\,dx'=
\left(\,\mathbf{v} \times \mathbf{j}_{K}  \,\right)\times
\mathbf{v}\,,\label{forcekj}\\
\mathbf{j}_{K}& =  \int K(\mathbf{r} - \mathbf{r'})\,
\mathbf{v}'\,f_{1}(x',t)\,dx' =
\int K(\mathbf{r} - \mathbf{r'})\,\mathbf{j}
(\mathbf{r}',t)\,d\mathbf{r}' \label{jkj}\,\,.
\end{align}

Using \eqs{basiceq} the equations of motion for the basic physical quantities can be derived. The obvious conservation law for the
number of particles is obtained from \eqs{basiceq} by integration
over $\mathbf{v}$:
\begin{equation}\label{eq_continuum}
	\frac{\partial \rho}{\partial t} + \divv{\mathbf{j}}= 0\,\,,
\end{equation}
where
\[\rho(\m{r},t) = n\,\int f_{1}(x,t)\,d\m{v}\,.\]
The equation for the hydrodynamic velocity field $\mathbf{u}$ is obtained from \eqs{basiceq} by multiplication by $\mathbf{v}$ and integrating over $\mathbf{v}$ with one-particle distribution function $f_{1}$:
\begin{equation}\label{eq_u}
\frac{\partial
\left(\, \rho\,u_i \,\right)
}{\partial t} +\frac{\partial }{\partial x_{j}}
\left(\, \rho\,u_i\,u_{j} \,\right) = -
\frac{\partial P_{ij}}{\partial x_{j}}+\mathcal{F}_i+
n\,\int \delta\,v_{i}\,I(x,t)\,d\mathbf{v}\,\,,
\end{equation}
where $\delta \m{v} = \m{v} - \m{u}$ so that
\[\int \delta v_{i}\,f_{1}\,dx = 0\,,\]
and the pressure
tensor $\mathbf{P}$ can be decomposed into
scalar and traceless components in a standard way: %
\begin{equation}\label{eq_pressuretens}
  P_{ij} = n\,\int \, \delta v_{i}\,\delta
v_{j}\,f_{1}\,d\mathbf{v}
  = p_0\,\delta_{ij} + \pi_{ij}\,\,.
\end{equation}
Here $\pi_{ij}$ is the traceless part which can be identified
with the viscosity stress tensor
\begin{equation}\label{eq_visctens}
  \pi_{ij} = n\,\int \,
\left(\, \delta v_{i}\,\delta v_{j} -  \f{\delta
\m{v}^2}{d}\delta_{ij}\,\,\right)
\,f_{1}\,d\mathbf{v}\,\,.
\end{equation}
The inner pressure is determined as following:
\begin{equation}\label{eq_pressure0}
  p_0 = n\,\int   \f{\left(\, \delta \mathbf{v} \,\right)^2}{2}\,
  f_{1}\,d\mathbf{v}= \f{\rho}{2}
\left(\, 1-\mathbf{u}^2\,\right)\,\,.
\end{equation}
Also
\begin{equation}\label{eq_forcemf}
  \boldsymbol{\mathcal{F}}(\mathbf{r},t) =
  n\,\int \mathbf{F}\,f_{1}(x,t)\,d\mathbf{v}\,\,,
\end{equation}
is the mean force density of the self-consistent field. In the hydrodynamic limit the spatial
scale of the variations for $f_1$ is much bigger than that of the
kernel $K(\m{r} - \m{r}')$. Therefore the microscopic kernel
should be put to $\delta$-function and $\m{j}_{K} \equiv \m{j}$.
Thus the force can be decomposed as following:
\begin{equation}\label{eq_force1}
  \boldsymbol{\mathcal{F}}(\mathbf{r},t) =
  p_0\,\mathbf{j} - \mathbf{j}.\boldsymbol{\pi}\,\,.
\end{equation}
Note that the authors of \cite{spp_degondmacrolimit_comptrend2007,spp_degondcontlimit_mmapp2008}  introduced the mean force which was proportional to the density gradient. But the representation \cite{spp_degondmacrolimit_comptrend2007,spp_degondcontlimit_mmapp2008}  did not provide clear distinction between this force and the dynamic term of the stress tensor because the authors used the  field of the unit director $\boldsymbol{\Omega} = \m{j}/|\m{j}|$ of the local flux.

Equation \eqref{basiceq} along with \eqref{collisint} is the basis for the construction of the hydrodynamic equations for the SPP fluid. The derivation of the proper form of the collision integral $I$ is out of the scope of the paper. Below we consider the simplified case of the ideal fluid where all correlation effects vanish ($I=0$, $\mathbf{\pi}=0$, {\it etc}). Nevertheless due to the presence of mean force it is possible to get the nontrivial dynamic of the SPP system.

\section{Hydrodynamic limit of ideal self-propelled Vicsek fluid}\label{sec_idealhydro}
The hydrodynamic limit of the ideal SPP-fluid corresponding to the SVM (an equation analogous to the Euler equation) can be obtained in a usual way neglecting in Eq.~\eqref{eq_u} all terms  generated by the correlations. These are the collision integral and the viscosity stress tensor $\boldsymbol{\pi}$ in Eq.~\eqref{eq_u}. Due to this the second term in the mean force \eqref{eq_force1} can be omitted when substituting it into Eq.~\eqref{eq_u} and Eq.~\eqref{eq_u} is simplified:
\begin{equation}\label{eq_ideal_der1}
\frac{\partial
\left(\, \rho\,u_i \,\right)
}{\partial t} +
\frac{\partial }{\partial x_{j}}
\left(\, \rho\,u_i\,u_{j} \,\right) = -
\frac{\partial p_0}{\partial x_{i}}+p_0 \rho u_i\,\,.
\end{equation}
Using Eq.~\eqref{eq_continuum} we finally obtain:
\begin{equation}\label{eq_ideal}
  \f{d\,\m{u}}{d\,t} = -\f{\nabla\,p_0}{\rho} + p_0\,\m{u}\,
\end{equation}
where $p_0$ is given by the Eq.~\eqref{eq_pressure0}. The
first term in the right-hand side reminds the corresponding term of the Euler equation for the Newtonian ideal fluid. The second term as it follows from Eq.~\eqref{eq_force1} is the mean force due to the neighbors. It has obviously aligning character and has the structure of Landau theory like terms introduced phenomenologically \cite{spp_tonertu_prl1995}. One can expect that taking into account the collision terms will renormalize the unit coefficients of the corresponding power terms.  The resulting Eq.~\eqref{eq_ideal} represents asymptotic form of the pressure terms in {\it zero} noise limit. In the kinetic approach \cite{spp_bertingregoire_jphysa2009} the corresponding term has the form $\left(\mu - \xi\,\rho^2\m{u}^2 \right)\,\rho\m{u}$ where $\mu,\xi$ are functions of density and the noise intensity. {Here we have $\mu=\xi \rho^2=\rho/2$.
In the next section it will be discussed, how these coefficients will change if the noise is taken into account.

The simplest solutions of Eq.~\eqref{eq_ideal} are the homogeneous ones for which both $\rho$ and $\m{u}$ do not depend on the spatial coordinates. It is easy to see that such solutions have the form:
\begin{equation}\label{homogsolut}
  \rho = \rho_{0}\,, \quad |\m{u}(t)| = \f{1}{\sqrt{1+e^{-\rho_0\,t}\,
\left(\, \f{1}{u^2_0} -1\,\right)\,,
  }}
\end{equation}
%
Though Eq.~\eqref{eom} has sense only in dimensions larger than 1 we think that it is instructive to consider 1-dimensional case assuming translational invariance along another directions. Besides, 1-dimensional models of the active particles are useful because of their simplicity, though they can be different from the more realistic models \cite{spp_activespins1dim_prl2013}. Also, recent experimental studies \cite{spp_activecolloid_nature2013} are in essential 1-dimensional.

Clearly, Eq.~\eqref{homogsolut} demonstrates the relaxation to the ordered state with the corresponding relaxation time $\tau_{rl} = 1/\rho_0$. Note that this fact is in an accordance with the result for the relaxation time of the velocity alignment in the mean-field approximation \cite{spp_kineticus_proc2009}.

It is natural to search for the running wave solutions
of the form $\rho(\m{r}-\m{v}\,t)\,,\m{u}(\m{r}-\m{v}\,t)$ using obtained equations.  We consider the simplified case of 1-dimensional geometry to get the exact solution. Here we give a simple example of 1D running wave solution $\rho(z)\,,u(z)$ with $z= x-v_0\,t$. The equations \eqref{eq_continuum} and \eqref{eq_u} for such solutions are:
\begin{align}
  - v_0\,\f{d \rho}{dz}+ \f{d \rho\,u}{dz} &= 0 \label{eqs_solitary1}\\
  - v_0\,\f{d u}{dz}+ u\,\f{d\,u}{dz} &= -\f{1}{\rho}\,\f{d}{dz}
\left(\, \f{\rho}{2}\left(\, 1-u^2 \,\right) \,\right)
+\f{u}{2}\,\rho\,
\left(\, 1-u^2 \,\right)\label{eqs_solitary2}\,.
\end{align}
This system can be integrated explicitly:
\begin{equation}\label{eq_rho1dimsolut}
\rho(z) = \f{C}{u(z) - v_0}\,,
\end{equation}
and
\begin{equation}\label{eq_v1dim}
\left(2 v_0^2-1\right) \log u-(v_0-1) v_0 \log (1-u)-v_0 (v_0+1) \log (u+1) = C\,z\,\,,
\end{equation}
where $C$ is the constant of integration and due to the translational invariance we omit the integration constant for $z$.

We search for the nonsingular solution $0<u(z)<1$ in entire interval $-\infty< z < +\infty$. That means that the range of values of the lhs of Eq.~\eqref{eq_v1dim} spans the interval $(-\infty, +\infty)$. Such situation occurs either if $v_0<-1/\sqrt{2}$ and $C>0$ or if $v_0>1$ and $C<0$ correspondingly. Yet the asymptotic behavior of $u(z)$ at $t\to \infty$ does not depend on sign of $v_0$. Indeed from Eq.~\eqref{eqs_solitary2} and \eqref{eq_rho1dimsolut} it follows that:
\begin{equation}\label{eq_intdvizh}
 \f{d}{d\,z}\left(\, u+\f{1}{2\,C}\rho\,
\left(\, 1-u^2 \,\right)
  \,\right)= \f{C}{2}
\f{\left(\, 1-u^2 \,\right)}{\left(\, u-v_0 \,\right)^2}\,u \sim {\rm sign}\, C\,.
\end{equation}
So that $u\to 1$ at $z\to +\infty$ if $C>0$ and $u\to 1$ at $z\to -\infty$ if $C<0$.

Consider the case $v_0<-1/\sqrt{2}$ (see Fig.~\ref{fig_spp1d_unst}). For such solutions the disordered region has \textit{higher} density than the ordered one. Then the fact that $u \to 1$ at $t\to \infty$ can be interpreted as the disappearance of
unstable state. This is expected result because of the consideration of the ideal regime of low noise where all collision terms are neglected. Fig.~\ref{fig_spp1d} shows the solution with $v_0>1$ and the density of the ordered state is greater than that of the disordered one. These solutions are similar to the switching wave solutions in active bistable media \cite{book_sinergloskutmikh}.They are not traveling bands observed in the numerical experiments \cite{spp_chaitesumul_pre2008,spp_bertingregoire_jphysa2009}, because we consider the system far away from transition point, in ordered regime, so the bands don't exist \cite{spp_chaitesumul_pre2008}.


 \begin{figure}
 \includegraphics[scale=0.5]{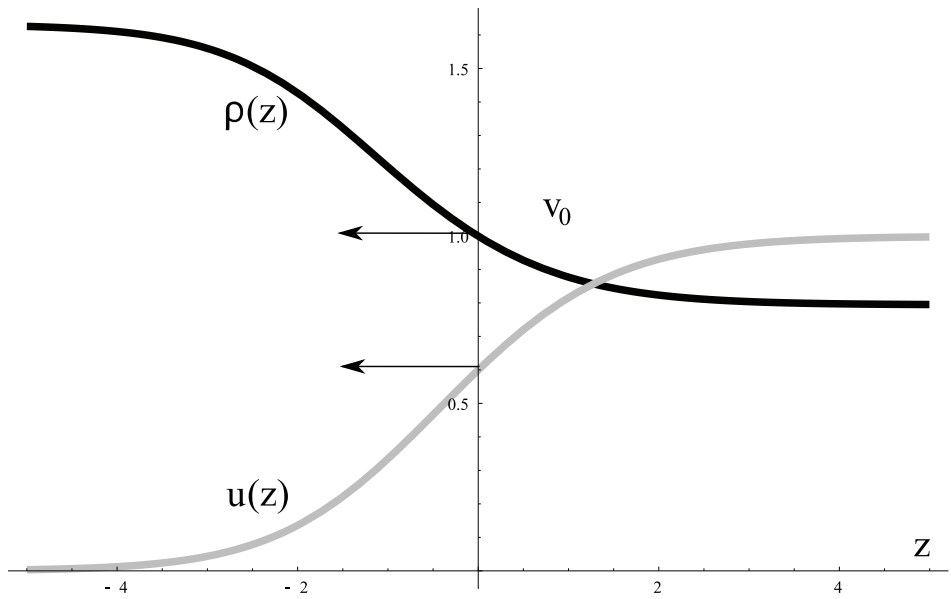}\\
   \caption{Density and velocity profiles with $v_0 =
 -1$.}\label{fig_spp1d_unst}
 \end{figure}

 \begin{figure}
 \includegraphics[scale=0.75]{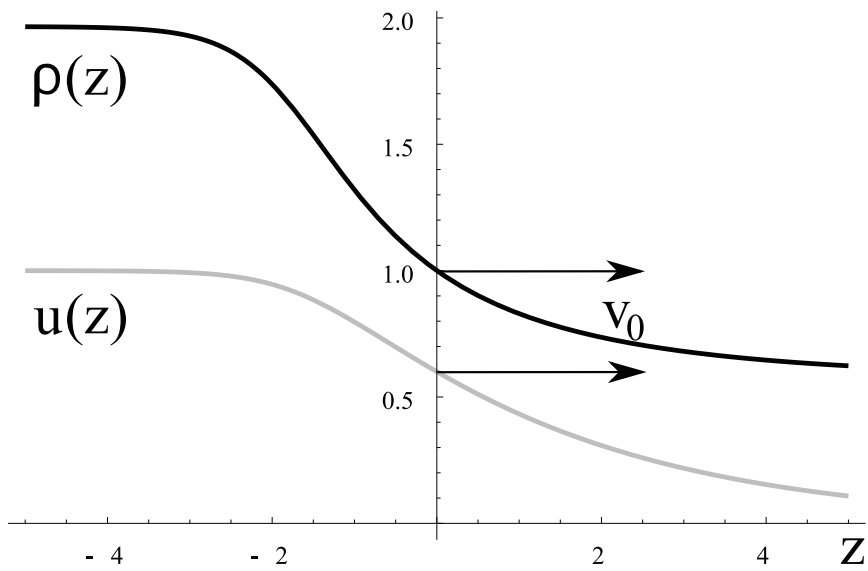}\\
   \caption{Density and velocity profiles with $v_0 =
 \sqrt{2}$.}\label{fig_spp1d}
 \end{figure}

In Fig.~\ref{fig_3d} we also show the solution of equations
\eqref{eq_continuum}, \eqref{eq_ideal} under different initial
states with periodic boundary conditions. These solutions also tend to the ordered state $u(x,t)\to 1$.

Detailed analytical study of more realistic 2D situation is left for future work.

Obtained results indicate that the ideal fluid approximation still retains the basic features of initial self-propelling system
despite neglecting the correlations. It shows how different
terms introduced from the heuristic arguments earlier (see e.g.
\cite{spp_tonertu_prl1995,spp_us_eurphyslet2005,spp_tonertu_prl1998}) may arise.

\begin{figure}
\subfigure[\,\,$u(x,0)=1-\cos^{8}
\left(\, \f{\pi\,x}{2L} \,\right)
\,,\,\, \rho(x,0) =
1+\cos^4{\left(\f{\pi\,x}{2L}\right)}$]{\includegraphics[scale=0.45]{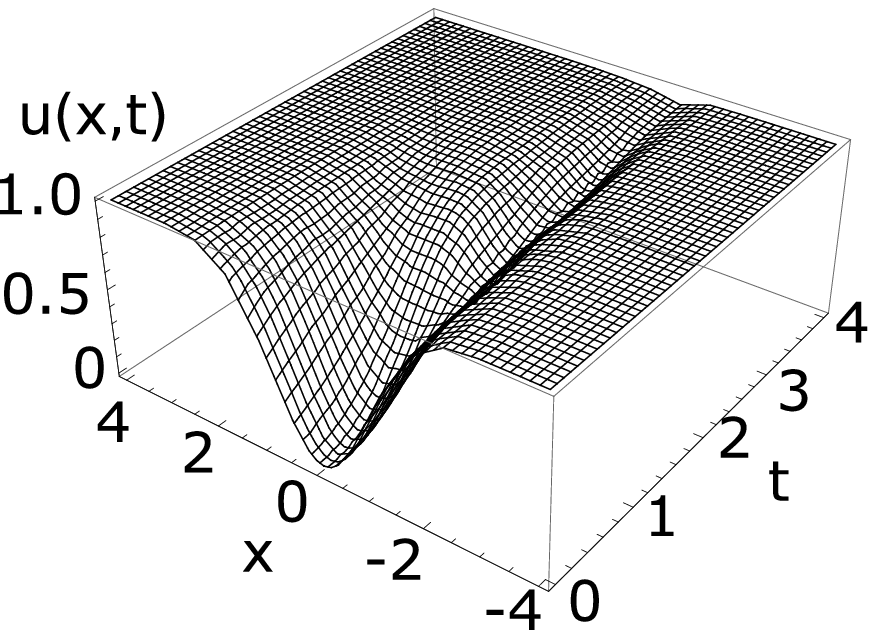}\,\,
\includegraphics[scale=0.45]{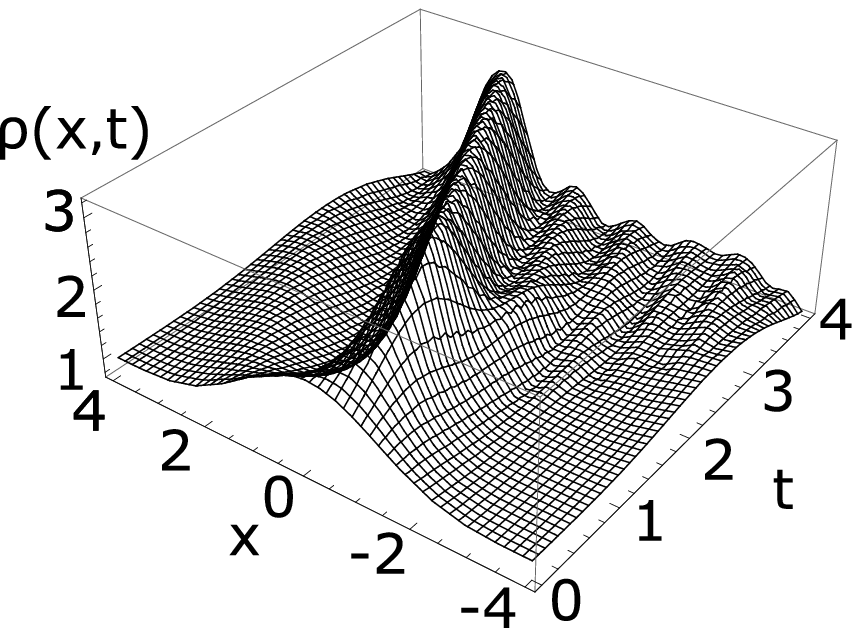}}
\subfigure[\,\,$u(x,0)=1/4\,,\,\, \rho(x,0) =
1+\cos^4{\left(\f{\pi\,x}{2L}\right)}$]{\includegraphics[scale=0.45]{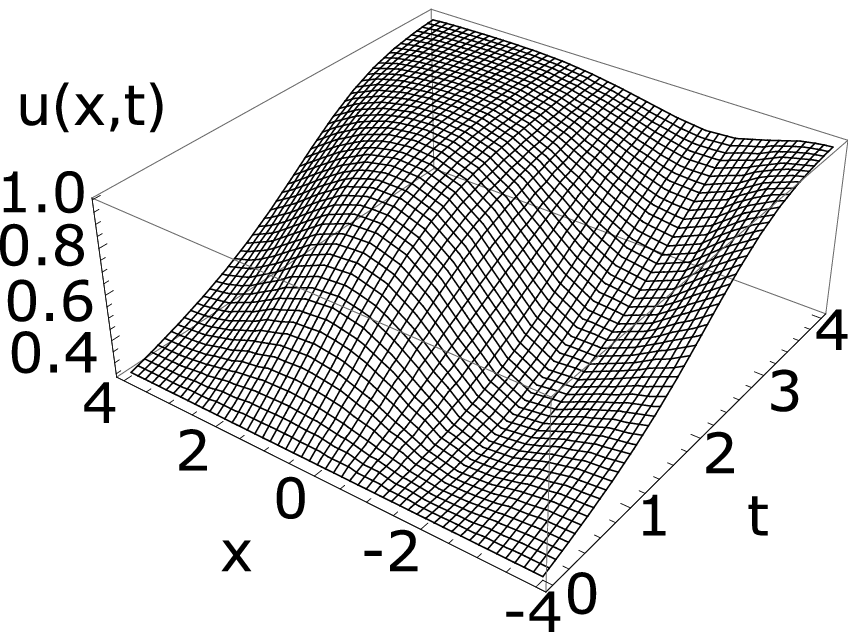}\,\,
\includegraphics[scale=0.45]{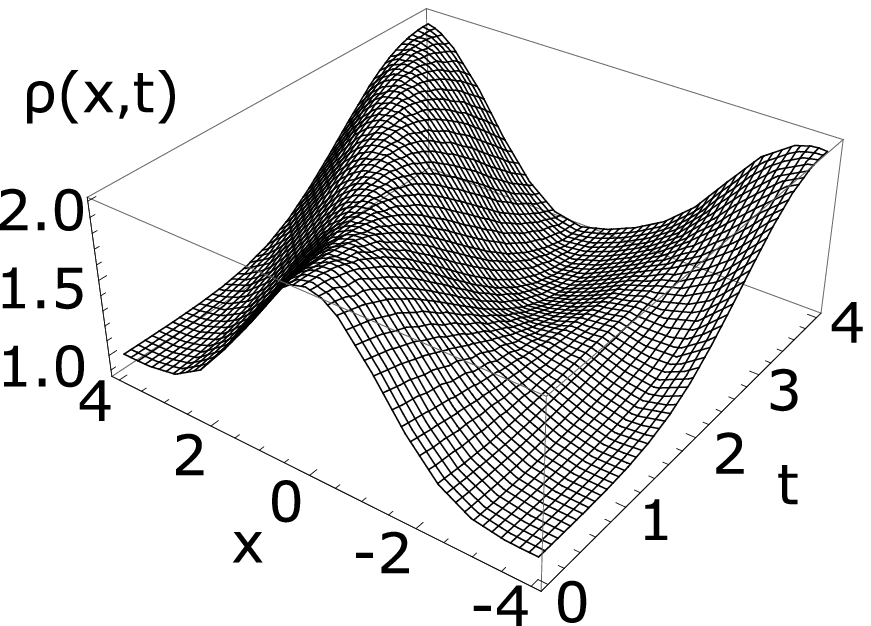}}
  \caption{Dynamics of the velocity $u(x,t)$ and density
$\rho(x,t)$ under different initial conditions. The video visualizations of these 3D plots are given in supplementary materials}\label{fig_3d}
\end{figure}

\section{Taking noise into account}
\label{sec_visc}

We have considered above the case of the ideal Vicsek fluid with the intention to derive an equation of Euler-kind for self-propelled fluid. But, in the very beginning of previous section we have mentioned that taking the noise into account may change the coefficients $\mu$ and $\xi$.
We are going to  go from {\it zero} noise limit to {\it low} noise limit where the system is far away from transition point and we can use {\it local equilibrium
approximation}.
In the local equilibrium
we use the distribution function of the equilibrium state of the system with local characteristics.

 Indeed, as it follows from
\cite{spp_kineticus_proc2009,spp_degondmacrolimit_comptrend2007,spp_cvalattic_pre1995} in the local equilibrium approximation for the ordered phase the distribution function can be approximated as von Mises-Fisher distribution \cite{spp_cvalattic_pre1995,spp_hydroDegondFrouvelleLiu_nonlin2013}:
\begin{equation}\label{eq_f1local}
f^{(0)}_{1}(\mathbf{r},\theta) = C e^{a \cos \theta}
\end{equation}
where $C$ and $a$ are some functions dependent on
the local value $\mathbf{u}(\mathbf{r})$ of the order parameter. The latter depends on the noise amplitude and the local density $\rho(\m{r})$. Another argument to use this form of the distribution function is that our hydrodynamical model in simple limit of Eq.~\eqref{homogsolut} shows the same relaxation properties as the model from \cite{spp_kineticus_proc2009}.
Obviously, Eq.~\eqref{eq_f1local} can serve as the interpolation between homogeneous distribution function $f = \f{1}{2\pi}$ in the disordered state and  the formal series representation of the distribution function near transition point to the ordered regime  \cite{spp_bertingregoire_jphysa2009}:
\[f_{1}(\mathbf{r}, \theta,t) = \f{\rho(\mathbf{r},t)}{n} \left( \f{1}{2\pi}+ \f{1}{\pi} u(\m{r},t)\,\cos\theta+ \ldots \right) \,\,\,.\]
We can specify explicitly $C$ and $a$  taking from \cite{spp_kineticus_proc2009}, and remembering that $u$ and $\rho$ depend on $r$:
\begin{equation}
f^{(0)}(\theta;r)  = \frac{1}{2 \pi I_0 (\lambda \frac{u}{D})}
\exp\left(\lambda \frac{u}{D} \cos \theta\right)\,\,,
\label{eq:distr}
\end{equation}
where $\lambda=\pi d_0^2 \rho$ is the average number of neighbors ($d_0$ is the radius of interaction), $D$ is the angular diffusion coefficient. $I_0$ is modified Bessel function of the first kind.

In such approximation the viscous tensor is:
\begin{equation}\label{eq_tensor_integral}
\pi_{ij}^{(0)} = \int\,
\left[ 	\delta v_i \delta v_j 	-
\delta_{ij}
\frac{(\delta \mathbf{v})^2}{2}\right] f^{(0)}
d\mathbf{v}
\end{equation}
The off-diagonal components  $i\neq j$ vanish
\begin{equation}
  \pi_{12} = \pi_{21}= \int \delta v_1 \delta v_2 \, C \, e^{a \cos \theta} d \theta = C \int (\cos\theta-u)\sin\theta e^{a\cos\theta}d\theta = 0
  \label{eq:pi12}
\end{equation}
according to the symmetry. This leads to the absence of the shear viscosity in the low noise limit. The diagonal part of the tensor
is determined by the quantity:
\begin{equation}
  \Pi^{(0)} =\frac{1}{2}\int [\delta v_1^2 - \delta v_2^2] f^{(0)}
    d \mathbf{v}
    =\frac{1}{2}\int [(\cos \theta - u)^2 - \sin^2\theta] f^{(0)}
    d \theta\,.
  \label{eq_pi11}
\end{equation}
which is not zero in general. $\Pi^{(0)}$ can be computed using Eq.~\eqref{eq:distr}:
\begin{equation}
\Pi^{(0)}=\frac{1}{2}
\left(
1 + u^2 - \frac{2 (D+\lambda u^2) I_1\left(\lambda \frac{u}{D} \right)}
{u \lambda I_0\left(\lambda \frac{u}{D} \right)}
\right)\,.
\label{eq:PI_full}
\end{equation}

We want to write explicitly Eq~\eqref{eq_u} (in 1D).
%
The series expansions of $\Pi^{(0)}$ and $\partial \Pi_0 / \partial x$ are:
\begin{equation}
\Pi^{(0)} = \frac{\alpha}{2}u^2
+ O(u^4)
\label{eq:Pi0_approx}
\end{equation}
and
\begin{equation}
\frac{\partial \Pi^{(0)}}{\partial x} =
\frac{\partial u}{\partial x}(\alpha u + \beta u^3)+O(u^5)\,\,,
\label{eq:Pi0dx_approx}
\end{equation}
where
$\alpha = 1 - \lambda/D+  \lambda^2/(8 D^2)$
and
$\beta = \lambda ^3 (6 D-\lambda ) / (24 D^4)$.

Now, the rewritten Eq.~\eqref{eq_u} is:

\begin{equation}
\frac{\partial u}{\partial t} =
-\frac{1}{2\rho} \frac{\partial \rho}{\partial x}(1-u^2)
+ \left(\frac{\rho}{2} - \frac{\alpha}{\rho} \frac{\partial u}{\partial x}\right) u - \left(\frac{\beta}{\rho}\frac{\partial u}{\partial x} +\frac{\rho}{2}+\frac{\alpha}{2}\right) u^3
\label{eq_u_rewritten}
\end{equation}

We can compare the coefficients that we have obtained with the corresponding ones from the work \cite{spp_bertingregoire_jphysa2009}.
So, in our approximation
\begin{equation}\label{eq:our_coeffs}
\mu = \frac{1}{2} \rho\,,\,\,\,\xi = \frac{1}{2\rho}+\frac{\alpha}{2 \rho^2}\,\,.
\end{equation}
In  \cite{spp_bertingregoire_jphysa2009} these coefficients are (if one assumes $\sigma \rightarrow 0$):
\begin{equation}
\label{eq:Bertin_coeffs}
\mu_{B} = \frac{8 \rho }{3 \pi } \approx 0.85 \rho\,,\,\,\xi_{B} = \frac{4}{\pi  \rho }-\frac{6 \sigma ^2}{\pi  \rho }+O\left(\sigma ^4\right)\,.
\end{equation}
where we put the radius of interaction $d_0 = 1$ and velocity $v_0=1$.

The correct comparison of these coefficients would be possible if for some interval of $\sigma$  $|\mathbf{u}|\rightarrow 1$. Unfortunately it doesn't exists for the model of \cite{spp_bertingregoire_jphysa2009}. Nevertheless,
from Eq.~\eqref{eq:our_coeffs} and Eq.~\eqref{eq:Bertin_coeffs} we see that $\mu$ and $\mu_{B}$ have the same dependence on $\rho$ but numerical factors differ. Density dependencies of $\xi$ and $\xi_{B}$ do not match exactly though both have terms inverse in density.

As a conclusion we can state that derived Euler-like equation for the Vicsek fluid \eqref{eq_ideal} can be considered as a zero-order approximation for more sophisticated hydrodynamical equations.




Obtained results show that in this respect SPP-fluid behaves differently from molecular liquids and even from the fluids formed by the active Brownian agents.
For the ordinary fluid the local equilibrium is given by the Maxwell distribution and $\pi_{ij}^{(0)} = 0$ due to the basic properties of the collision integral \cite{book_klimontovich_statphys}.
We have shown that the shear viscosity is absent in the Vicsek-like fluid in local equilibrium approximation given by Eq.~\eqref{eq_f1local}, where the system is close to be ordered and acting noise is small but non-zero. We think that this result is a consequence of alignment rule as it reduces the disorder in velocity distribution. Indeed, the interaction of Viscek type in this respect is drastically different from molecular collisions which give rise to the entropy production and therefore to the dissipation. Sure the detailed analysis of the collision terms is needed in order to clarify this question.
Previous studies showed that the viscosity term appeared in the hydrodynamic equations, and was present even in the limit of zero noise. The presence of this term could be an interesting question of discussion of the limits of the assumptions done in the studies. It should be noted that the equation of ``Vicsek hydrodynamics'' of Degond and coll. \cite{spp_degondmacrolimit_comptrend2007,spp_degondcontlimit_mmapp2008}  does not contain Navier-Stokes viscosity term.

So, this question needs to be resolved. The simplest way to study the propertiese (including existance) of viscosity is a numerical simulation. To do this one can perform simulations of self-propelled particles system subjected to shear (as example \cite{spp_gonnellashear_cejp2012}) and study the profile of the velocity field $\mathbf{u}$ that appears in such system. In the usual viscous liquid (Couette flow) linear profile (gradient) appears. The equations presented in, for example, \cite{spp_bertingregoire_jphysa2009} lead to a non-trivial and non-linear profile due to nonlinear terms present along with the shear viscosity term. Expected zero viscosity should result in the absence of the nontrivial profile of the order parameter, except, probably, very thin layer close to the aligning boundary, caused by the specific mechanism of interaction. This numerical experiment can explain the difference in the discussed approaches.  We have performed a first simple check for a small system and preliminary results show the absence of the order parameter profile \cite{spp_posterus_iwnet2012}. More accurate studies are needed to answer the question in full details.

\section{Discussion}
In this paper we use the standard method of microscopic phase density to derive the general form of the kinetic equation for the Vicsek-like model. We obtain the form of the collision terms which in principle gives the possibility to obtain the closed kinetic equation for one-particle distribution function $f_1$.
Such an approach is more closely connected with the equation of motion and could be easily modified to include different kind of velocity constraints and noise perturbation to dynamics.
We have considered the simplest mean-field approximation and checked that the corresponding hydrodynamic equations have solutions displaying the transition to the ordered state.
The obtained equation \eqref{eq_ideal} plays a role of the Euler equation for the self-propelled fluid of the Vicsek type.
Its homogeneous solution shows the same time relaxation of alignment as that obtained in the mean-field kinetic approach \cite{spp_kineticus_proc2009}.

The comparison of the coefficients from our equations with well-known model \cite{spp_bertingregoire_jphysa2009} shows some similarities, though our model took the noise into account only in mean-field limit. Our model has the possibility of extending to more complete description of collisions and correlations, which will lead to more accurate coefficients in our later work.
Note that we considered the rule of alignment to the local flux of the neighbors, but not to the average velocity as in original Vicsek model. This simplifies the calculations because of the additivity and collision integrals depend on pair correlation function only. Although this simplification seems to be appropriate from the physical point of view,
the question of mathematical equivalence between these two ways remains open and proper numerical study is required. Usage of the original Vicsek rule of alignment to the average velocity in principle leads to essentially many-particle collision terms.
Thus in such an approach it is possible to consider in unified way both the kinetic equation obtained in Boltzmann-like derivation scheme  \cite{spp_gregorie_kinetic_pre2006, spp_bertingregoire_jphysa2009} and master equation approach \cite{spp_kinetichydro_pre2011} which lead to essentially many-particle collision terms.

Also we have considered important question about the viscosity of the Vicsek-like model.
The calculation of the viscous tensor in the local approximation shows vanishing of the diagonal components of the tensor. We have checked this by looking at the velocity profile in a small test system. However proper numerical proofs of this result are required. Based on the previous studies \cite{spp_cvalattic_pre1995,spp_kineticus_proc2009} the form of distribution function has the Boltzmann-like form. Here we should note that the existence of shear viscosity and even non-Newtonian rheological properties \cite{spp_viscnonewton_prl2013} are expected for the dynamic models of active Brownian particles and for the modifications of Vicsek Model augmented with the repulsive potential forces \cite{spp_chaitesumul_pre2008} due to collisions. In such cases obviously the general framework of \cite{spp_tonertu_pre1998} is relevant phenomenology. As to the question about pure kinematic or nonholonomic models like SVM this question should be the subject of both accurate numerical experiment and peer theoretical analysis.
These and other questions will be considered in separate work.

\section*{Acknowledgments}
Authors are grateful to Professor Fernando Peruani and Dr. Francisco J. Sevilla for the fruitful discussion of the results.
V.K. thanks Mr. Konstantin Yun for support.
%
%
%
%

\section{References}
\bibliographystyle{elsarticle-num}


\end{document}